\documentclass[pra,floatfix,twocolumn,superscriptaddress,showpacs,preprintnumbers,longbibliography,nofootinbib]{revtex4-1}

\usepackage[utf8]{inputenc} 
\usepackage{bibunits} 
\usepackage{amssymb,amsfonts,amsmath}
\usepackage{float} 
\usepackage{amsthm}
\usepackage{enumitem}
\usepackage{bbold}
\usepackage[normalem]{ulem}
\usepackage{hyperref} 
\hypersetup{colorlinks=true, urlcolor=blue, citecolor=blue,linkcolor=blue}
\usepackage[all]{hypcap} 

\usepackage[mathscr]{euscript}
\usepackage{graphicx}
\usepackage{dcolumn}
\usepackage{bm}
\usepackage{psfrag}
\usepackage{epsfig}
\usepackage{multirow}
\usepackage{color}
\usepackage{bbm}
\usepackage[FIGTOPCAP,raggedright,nooneline]{subfigure}
\usepackage{verbatim}
\usepackage{upgreek} 
\usepackage{physics}
\usepackage{tikz} 

\newcommand{\T}[1]{\text{#1}}

\newcommand{\ignore}[1]{}

\newcommand{\eq}{Eq.\,}

\newcommand{\fig}{Fig.\,}

\newcommand{\rref} {Ref.\,}

\newcommand{\PT}{$\mathcal{PT}$}

\begin{document}

		\title{Quantum correlations in dissipative gain-loss systems across exceptional points}

		\author{Federico Roccati}
		\affiliation{Department of Physics and Materials Science, University of Luxembourg, L-1511 Luxembourg}
		
		\author{Archak Purkayastha}
		\affiliation{School of Physics, Trinity College Dublin, College Green, Dublin 2, Ireland}
		\affiliation{Centre for complex quantum systems, Aarhus University, Nordre Ringgade 1, 8000 Aarhus C, Denmark}
		
		\author{G.~Massimo Palma}
		\affiliation{Universit\`a  degli Studi di Palermo, Dipartimento di Fisica e Chimica -- Emilio Segr\`e, via Archirafi 36, I-90123 Palermo, Italy}
		\affiliation{NEST, Istituto Nanoscienze-CNR, Piazza S. Silvestro 12, 56127 Pisa, Italy}
		
		\author{Francesco Ciccarello}
		\affiliation{Universit\`a  degli Studi di Palermo, Dipartimento di Fisica e Chimica -- Emilio Segr\`e, via Archirafi 36, I-90123 Palermo, Italy}
		\affiliation{NEST, Istituto Nanoscienze-CNR, Piazza S. Silvestro 12, 56127 Pisa, Italy}
	
		\date{\today}
	
		\begin{abstract}
			We investigate the behavior of correlations dynamics in  a \textit{dissipative} gain-loss system.
			First, we consider 
			a setup made of two coupled lossy oscillators, with one of them subject to a local gain. This provides a more realistic platform to implement parity-time (\PT) symmetry   circumventing the implementation of a pure gain. We show how the qualitative dynamics of correlations resembles that for a pure-gain-loss setup. The major quantitative effect 
			is that quantum correlations are reduced, while total ones are enhanced. 
			Second, we study the behavior of these correlations across an exceptional point (EP) outside of the \PT-symmetric regime of parameters, observing how
			different behaviors across the EP occur only in the transient dynamics. This shows how \PT~symmetry plays a relevant role at large times.
			
		\end{abstract}
	
		\maketitle

		\section{Introduction}

		Parity-Time (\PT) symmetric Hamiltonians have attracted growing attention over the last two decades as they may possess a real spectrum despite being non-Hermitian (NH)~\cite{benderPRL1998}. A Hamiltonian $H$, not necessarily Hermitian, is  \PT-symmetric if it commutes with the anti-Hermitian operator \PT, where $\mathcal{P}$ and $\mathcal{T}$ are the parity and time reversal operators, respectively.
		
		A \PT-symmetric Hamiltonian may possess a real spectrum, in which case the \PT~symmetry is \textit{unbroken}, or a complex spectrum in the  \PT~\textit{broken} phase. These two regimes are separated by a so called 
		\textit{exceptional point} (EP). These are peculiar NH degeneracies as at an EP both eigenvalues \textit{and} eigenstates coincide, which are generally exhibited by NH Hamiltonians (even those lacking \PT~symmetry).
		
		The importance of \PT-symmetric Hamiltonians has grown also due to their experimental realization in optical platforms, where staggered real/complex refracting indices allow the implementation of classical \PT-symmetric potentials~\cite{ruterNP2010,regensburgerN2012,pengNP2014}.
		A paradigm of \PT-symmetric setups  is a pair of systems coherently exchanging energy, where one of which is subject to some form of leakage (loss) and the other one to pumping (gain)~\cite{el-ganainyNP2018,dastPRA2014,benderRPP2007}. 
		Gain-loss \PT-symmetric Hamiltonians often come up as effective tools to describe mean field dynamics~\cite{roccatiQST2021,roccati2022non} or quantum trajectories without jumps~\cite{MingantiPRA2019}.
		
		The realization of a pure gain is a major experimental hindrance in implementing  \PT-symmetric systems at the quantum level~\cite{scheelEPL2018}.

		This is still  an open problem, and many efforts have been recently made in the direction of characterizing the behavior of quantum features such as, e.g., correlations~\cite{KumarPRA2022}, transport~\cite{HuangNANOPH2017} and sensing~\cite{zhangAQ2018} near EPs~\cite{chen2021quantum,naghiloo2019quantum,QuirozPR2019,PurkayasthaPRR2020,caoPRL2020}. 
		
		In this work we study  quantum correlations (QCs) in a generalized gain-loss \PT-symmetric 
		system, emphasizing the differences with a pure gain-loss setup.

\begin{figure}
			\centering
			\includegraphics[width=0.8\columnwidth]{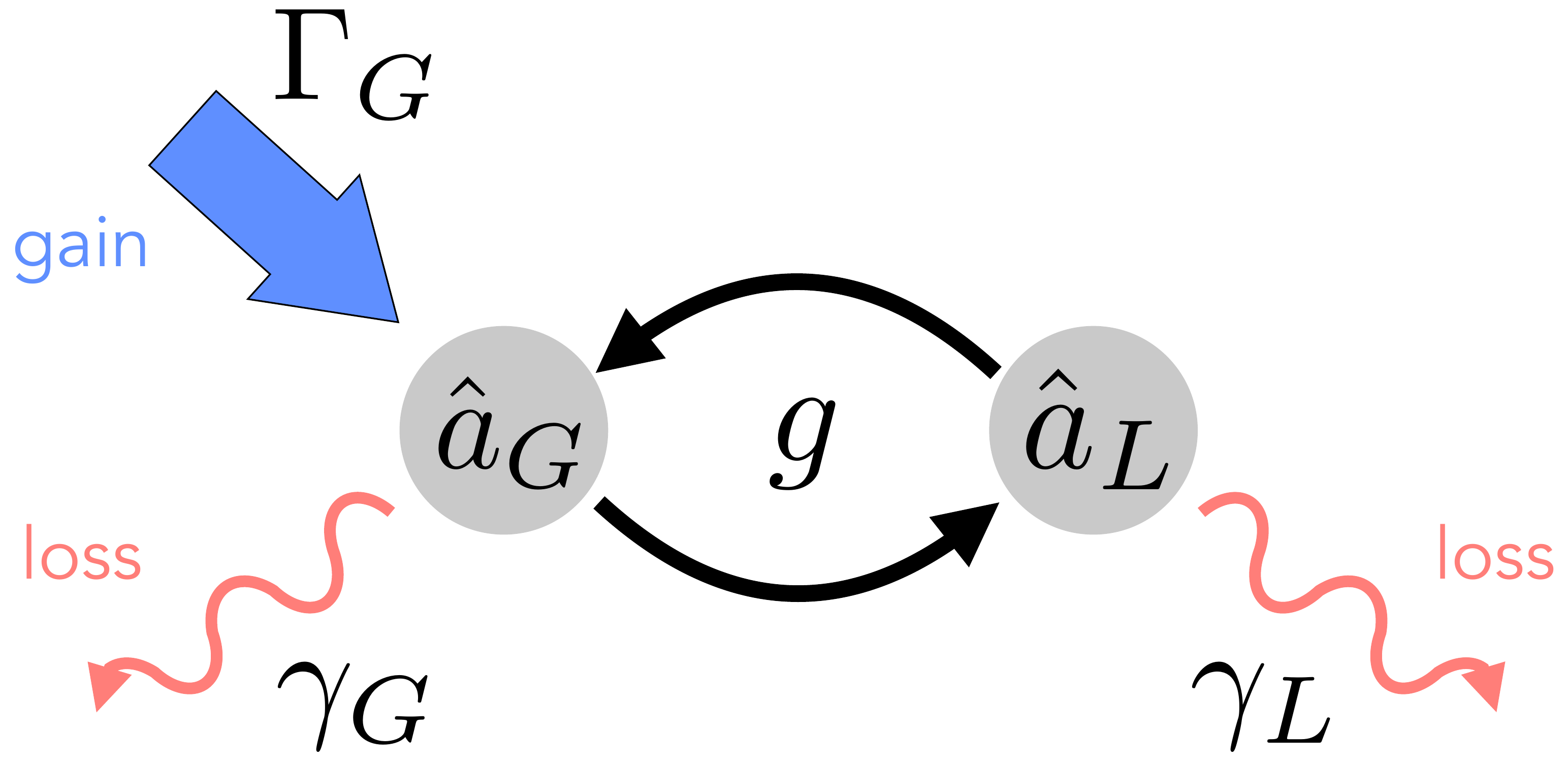}
			\caption{
				{\it Dissipative gain-loss system}.
				{A pair of quantum  oscillators $G$ and $L$ that coherently exchange excitations with rate $g$. Both modes $G$ and $L$ are subject to local losses with rates $\gamma_G$ and $\gamma_L$, respectively. Additionally, mode $G$ is subject to a local gain with rate $\Gamma_G$. If parameters are set so as to match the condition $\Gamma_G-\gamma_G=\gamma_L$, then the mean-field dynamics is described by a \PT-symmetric Hamiltonian.}}
			\label{setup_archack} 
		\end{figure}

		We first introduce our model, which we dub \textit{dissipative} gain-loss system (see Fig.~\ref{setup_archack}): a pair of coherently-coupled lossy modes with only one of these subject to gain. 
		Second, we study the dynamics of total and quantum correlations when the system is in its \PT-symmetric configuration, exhibiting critical behavior at the EP. Finally, we study the same correlations across another EP without \PT~symmetry, showing that different behaviors across it appear only in transient dynamics.

		\section{Dissipative gain-loss system}
		We consider a setup made of two quantum harmonic oscillators labeled by $G$ and $L$  whose closed  dynamics is 
		governed by the Hamiltonian
		\begin{equation}
		\hat H = g\,( \hat a_{L }^\dagger \hat a_{G}+{\rm H.c.})\,.
		\end{equation}
		This describes an excitation exchange at rate $g$ between the modes $\hat{a}_G$ and $\hat{a}_L$, that  are usual bosonic ladder operators.
		
		We consider now the  additional introduction of a local loss on either mode and a gain only on mode $L$. This makes the dynamics of the modes non-unitary and described 
		 by the Lindblad master equation~\cite{breuer2007,CICCARELLO20221} (we set $\hbar=1$ throughout)
		\begin{eqnarray}\label{newME}
		\dot\rho  
		& = &  - i[\hat H ,\rho] 
		+2\gamma_{L}\, \mathscr D[\hat a_L]\rho 
		\nonumber\\
		&    &
		+2\gamma_{G} \,\mathscr D[\hat a_G]\rho 
		+2\Gamma_G\, \mathscr D[\hat a_G^\dag]\rho 
		\end{eqnarray}
		with 
		\begin{equation}
		\mathscr D[\hat A]\rho=\hat A \rho \hat A^\dagger - \tfrac{1}{2}(\hat A^\dagger \hat A \rho+\rho\hat A^\dagger\hat A).\nonumber 
		\end{equation}
		Here, $\rho$ is the joint density matrix of the two modes. We work in a rotating frame  
		so as to eliminate the free Hamiltonian dynamics, which in turn does not affect correlations between the two modes.
		
		In the first line of Eq.~\eqref{newME} the second term describes the leakage of the mode $L$ at rate $\gamma_L$ into a local zero-temperature environment. In the second line of Eq.~\eqref{newME}, unlike \rref~\cite{roccatiQST2021}, the  first and second terms describe leakage and incoherent pumping on mode $G$ at rates $\gamma_G$ and $\Gamma_G$, respectively. 
		
		This configuration avoids the implementation of a pure gain (i.e., an incoherent pump), which is experimentally demanding~\cite{scheelEPL2018}. We refer to this as a \textit{dissipative} gain-loss system as, on top of gain ($\Gamma_G$) and loss ($\gamma_L$), we consider additional dissipation on mode $G$ at rate $\gamma_{G}$. If $\gamma_{G}=0$, the \textit{pure} gain-loss configuration is retrieved.

		\subsection{First-moment dynamics}

		Master equation~\eqref{newME} implies a Schr\"odinger-like equation $i \dot \Psi=\mathcal H\, \Psi$ for the mean-field vector $\Psi=(\langle \hat{a}_L\rangle, \langle \hat{a}_G\rangle)^T$ with
		\begin{equation}\label{newevolMeanVal}
		\mathcal H=\left(
		\begin{matrix} 
		-i\gamma_{L } & g \\
		g & i\, \tilde\Gamma_{G}
		\end{matrix} 
		\right)
		\end{equation}
		where $\tilde\Gamma_{G}=\Gamma_{G} -\gamma_G$ is the \textit{effective} gain rate and $\langle \hat a_{n}\rangle={\rm Tr}\left(\hat a_{n}\rho\right)$ for $n=L,G$.
		The corresponding dynamics is that of a classical dynamical system with \textit{unbalanced} gain and loss~\cite{bender2019pt}. 
		
		Matrix $\cal H$ has two complex eigenvalues given by
		\begin{equation}
		\mathcal E_\pm
		=
		-i\tfrac{\gamma_L-\tilde\Gamma_{G}}{2}
		\pm \sqrt{g^2-\left(\tfrac{\gamma_L+\tilde\Gamma_{G}}{2}\right)^2}\,,
		\end{equation}
		with associated non-orthogonal eigenstates.
		The former are purely imaginary (complex) if $\gamma_L+\tilde\Gamma_{G}>2g$ ($\gamma_L+\tilde\Gamma_{G}<2g$),
		coalescing at the EP $\gamma_L+\tilde\Gamma_{G}=2g$ where the corresponding eigenstates become parallel~\cite{el-ganainyNP2018}.

		\subsection{Second-moment dynamics}
		
		In order to go beyond a mean-field description of the dynamics, we consider the quantum uncertainties of the two modes described by the covariance matrix, whose entries are defined as
		$\sigma_{ij} = \langle \{\hat A_i, \hat   A_j\} \rangle -2 \langle \hat A_i \rangle \langle \hat A_j \rangle$,
		where $\hat A_i=( \hat a_L, \hat a_G,\hat a_L ^\dagger, \hat a_G^\dagger)$~\cite{gardiner2004}.

		Master equation~\eqref{newME} implies the following Lyapunov evolution equation~\cite{purkayastha2022lyapunov} for the  covariance matrix
		\begin{equation}\label{eqforsigmaY}
		\dot{\sigma} = Y\sigma +\sigma\, Y^{T} + 4 D 
		\end{equation}
		with
\begin{equation}\label{Ymatrix}
Y=\left(
\begin{matrix}
-i\mathcal H  & \textbf{0} \\
\textbf{0} & i\mathcal H^\dagger 
\end{matrix} 
\right)
\end{equation}
		and
        $D=\mathbb{1}_2\otimes\,{\rm diag}(\gamma_{L},\tilde\Gamma_{G}+2\gamma_G)/2$.
		This shows how the full dynamics indeed depends only the mean-field \PT-symmetric Hamiltonian $\mathcal H$.
		
		The Lindbladian~\eqref{newME} is quadratic. As such, it transforms Gaussian states into Gaussian states~\cite{ferraro2005}. Accorddingly, we will consider the  most classical \textit{uncorrelated} Gaussian initial state $\rho_0=\dyad{\alpha_L}\otimes\dyad{\alpha_G}$ with $\ket{\alpha_n}$ (for $n=L,G$) a coherent state of amplitude $\alpha_{n}$.
		The corresponding covariance matrix is $\sigma_0=\mathbb{1}_2\otimes\mathbb{1}_2$, that is the same of the vacuum state (regardless of amplitudes $\alpha_n$). This is because a coherent state $\ket{\alpha}$ contains only the noise of the vacuum, that is
		\begin{equation}
		\bra{0}\hat A ^2\ket{0}=\bra{\alpha}\hat A ^2\ket{\alpha} - \bra{\alpha}\hat A \ket{\alpha}^2
		\end{equation}
		with $\hat A = \hat x, \hat p$, for any amplitude  $\alpha$.
		
		Therefore, up to a shift $\Psi$, the joint state of the system is completely described  by the covariance matrix $\sigma$~\cite{ferraro2005}.
		One of the advantages of working with Gaussian states is that entropies, total and quantum correlations can be calculated as functions of the covariance matrix, see Appendix~\ref{appendix}.

		\begin{figure}
	\includegraphics[width=\linewidth]{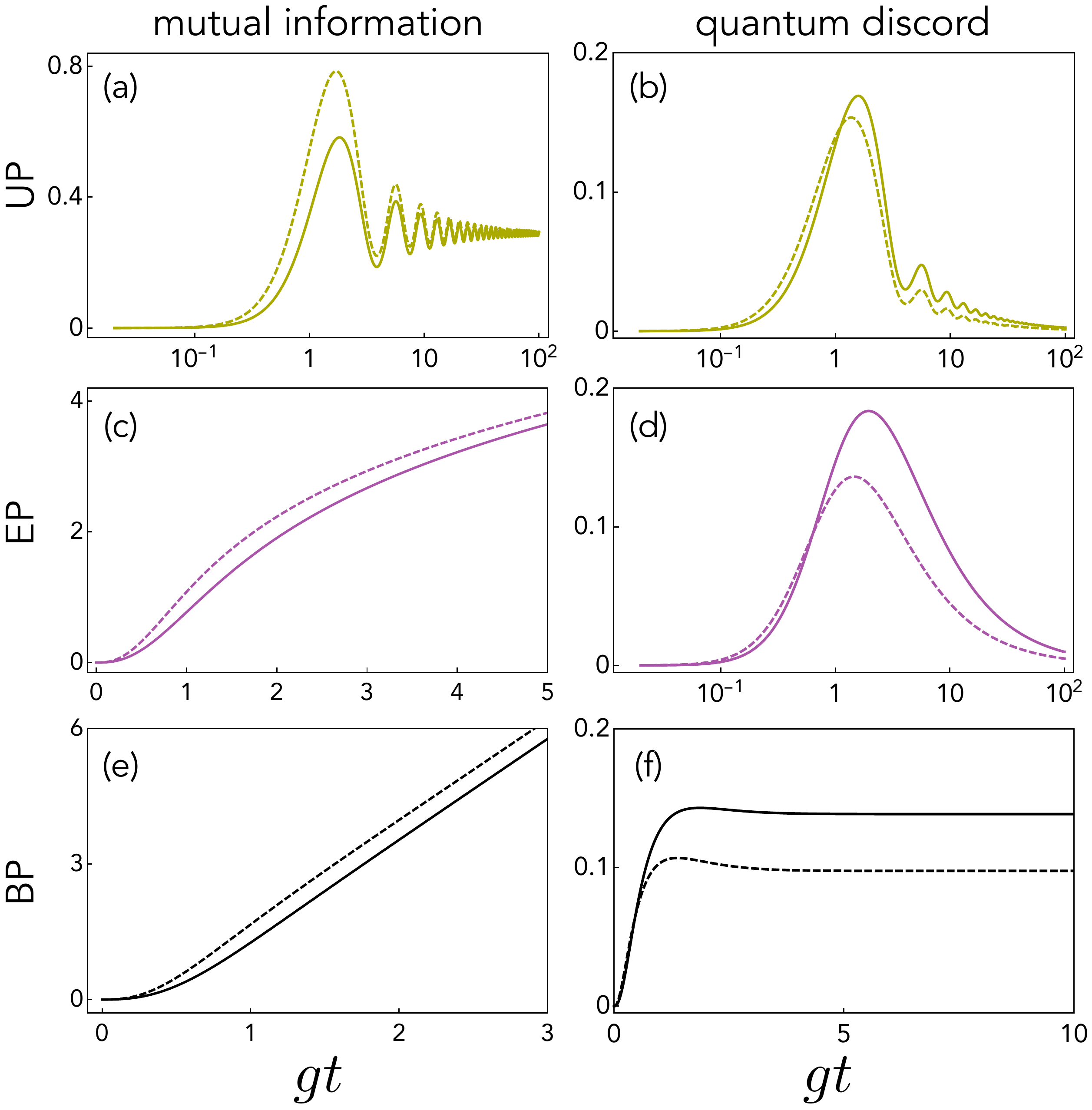}
	\caption{
		{\it Total and quantum correlations dynamics in the \PT~regime}.
		Time behavior of mutual information (left panels) and quantum discord (right panels)  in the unbroken phase (UP) [(a)-(b)], exceptional point [(c)-(d)] and broken phase (BP) [(e)-(f)].
		Solid (dashed): pure (dissipative) gain-loss configuration $\gamma_G=0\, (\Gamma_G/2)$, 
		with UP, EP and BP corresponding to $\Gamma_G=g/2,\, g,\, 3g/2\, (g,\, 2g,\, 3g)$, 
		respectively. 
	} \label{fig2}
\end{figure}

		\section{\PT-symmetric regime }
		
		The system parameters can be tuned so to implement  \PT~symmetry by balancing  loss and effective gain, $\gamma_{L }=\tilde\Gamma_{G}$, a condition which we will assume in this section.
		The eigenvalues now simply read 
		\begin{equation}
			\mathcal E_\pm
			=
			\pm\sqrt{g^2-\tilde \Gamma_G^2}\,\,.
		\end{equation}
		Accordingly, \PT~symmetry is unbroken (broken) for $\tilde \Gamma_G<g$ ($\tilde \Gamma_G>g$). Note that, since all rates are  positive, the \PT-symmetric  condition $\gamma_{L }=\tilde\Gamma_{G}$ imposes the further constraint  $\gamma_{G} <\Gamma_G$. Therefore the local bath connected to oscillator $G$ cannot be interpreted as a thermal environment since the detailed balance condition would imply $\gamma_{G} >\Gamma_G$~\cite{CICCARELLO20221,binder2019thermodynamics,De_Chiara_2018}.
		
		As shown in \fig\ref{fig2}, the dynamics of total and quantum correlations (see Appendix~\ref{appendix} for details) is qualitatively analogous to that of a pure gain-loss setup~\cite{roccatiQST2021}. 
		Mutual information, capturing the total amount of correlations (both classical and quantum) reaches a finite stationary value in the  unbroken regime $\tilde\Gamma_{G}<g$, linearly diverges  in the broken one $\tilde\Gamma_{G}>g$, and logarithmically diverges at the EP $\tilde\Gamma_{G}=g$ [see \fig\ref{fig2}(a)-(c)-(e)].
		On the other hand, quantum discord (measuring quantum correlations) vanishes in the unbroken phase and at the EP, though with different scalings, while it approaches a stationary value in the broken phase [c.f.~Fig.~\ref{fig2}(b)-(d)-(f)].
		\begin{figure}
			\includegraphics[width=0.95\linewidth]{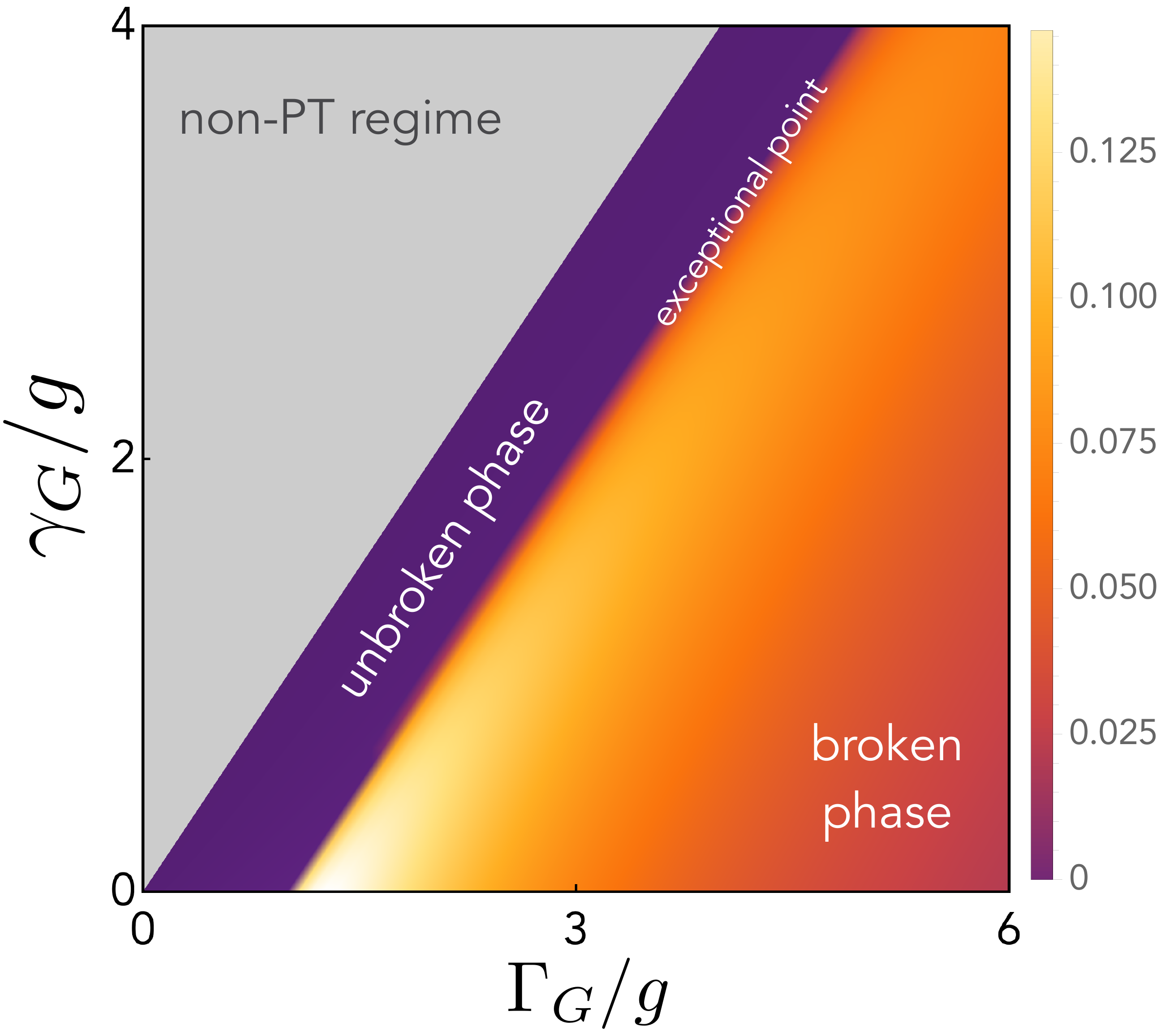}
			\caption{
				{\it Asymptotic quantum correlations  in the \PT~regime}. 
				Long-time QCs measured by quantum discord $\mathcal D_{LG}$ ($\mathcal D_{GL}$ displays  equivalent qualitative behavior). The \PT-symmetric regime, achieved by balancing \textit{loss} and \textit{effective gain}, corresponds to the region $\gamma_G>\Gamma_G$ (purple and orange). In this regime, for each $(\Gamma_G,\gamma_G)$, the value of $\gamma_L$ is chosen so that \PT~symmetry holds. In the unbroken phase, i.e., $\Gamma_G-\gamma_G<g$, quantum discord vanishes for any dissipation strength $\gamma_G$. Contrarily, in the broken phase it approaches a finite value, which decreases with $\gamma_G$.
			} \label{fig4}
		\end{figure}		
	
		It turns out that adding dissipation on mode $G$ is that to increase the total amount of correlations and eventually reduce quantum discord, as shown in Fig.~\ref{fig2}. This effect occurs as the gain process generally introduces mixedness into the system, which helps creating discord, while pure dissipation tends to enhance purity~\cite{roccatiQST2021}.

		Fig.~\ref{fig4} shows the behavior of asymptotic quantum discord in the available \PT-symmetric configurations ($\gamma_G<\Gamma_G$) in the parameter space. In the broken phase, quantum discord is finite and decreasing with $\gamma_G$. In the non-\PT-symmetric regime ($\gamma_G>\Gamma_G$), perfect gain loss balance cannot be achieved. Notwithstanding, the effective mean-field Hamiltonian exhibits EPs and a similar investigation can be conducted.

		\section{Correlations dynamics in the non-$\mathcal{PT}$ regime}
		
		The mean-field Hamiltonian for our master equation exhibits EPs even away from the \PT-symmetric regime. It is natural to wonder whether correlations are sensitive to these points.
		
\begin{figure}
			\includegraphics[width=0.9\linewidth]{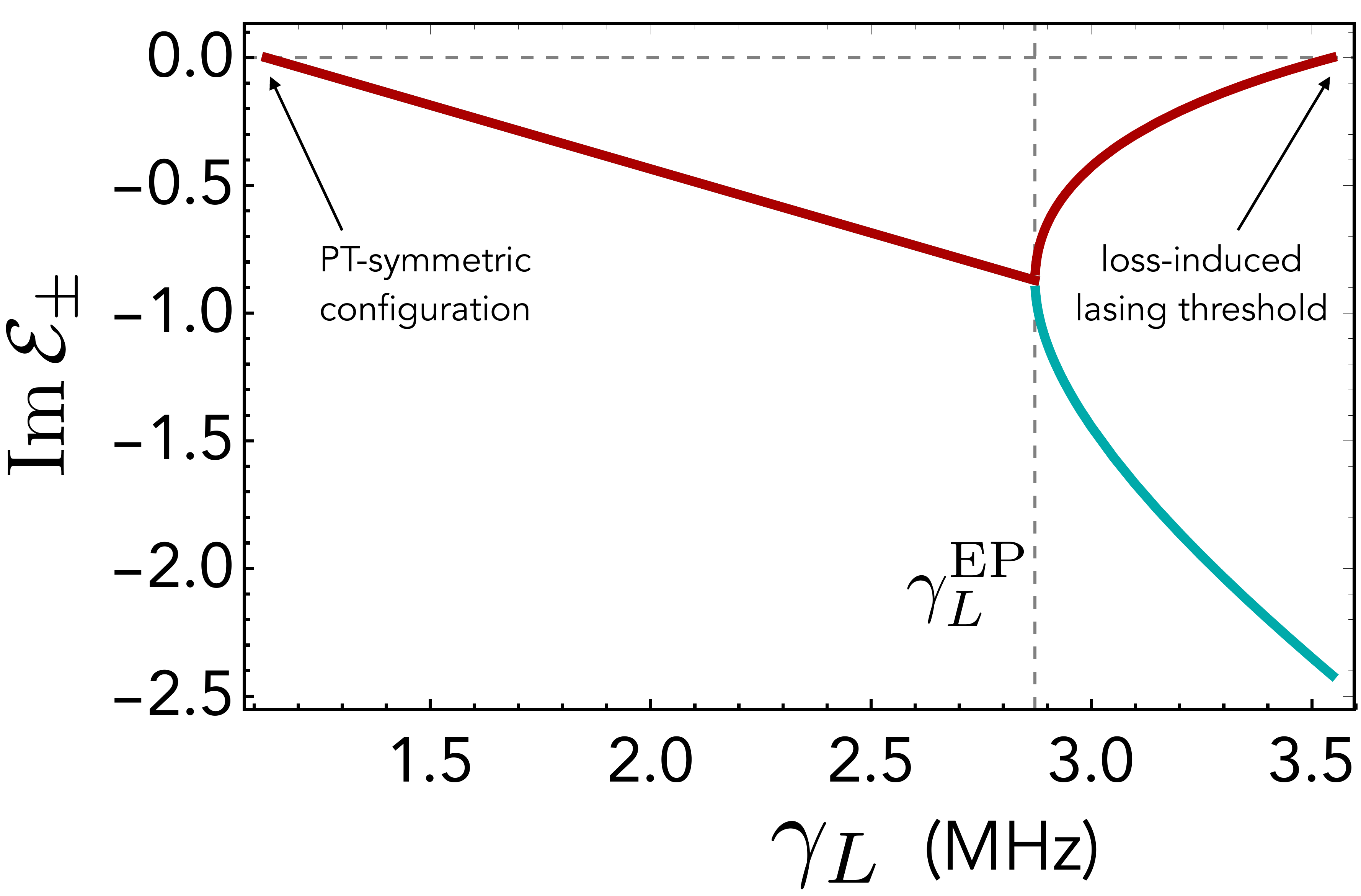}
			\caption{
				{\it Mean-field spectrum}. 
				Imaginary part of the eigenvalues of $\mathcal H$ as a function of the only-loss cavity dissipation rate $\gamma_L$.
			} \label{fig6}
		\end{figure}

		An EP arises when varying the loss strength $\gamma_L$  of the only-loss cavity ($L$) between the \PT-symmetric threshold 
		\begin{equation}
		\gamma_L^\T{PT}=\Gamma_G-\gamma_G\,,
		\end{equation}
		and the so called loss-induced lasing threshold~\cite{PurkayasthaPRR2020}
		\begin{equation}
			\gamma_L^\T{th} = 
			\frac{2g^2}{\Gamma_G - \gamma_G}\,.
		\end{equation}
		The EP occurs at
		\begin{equation}
			\gamma_L^\T{EP}= 2g
			+\gamma_G - \Gamma_G\,,
		\end{equation}
	    as shown in Fig.~\ref{fig6}. 
	    
	    Master equation~\eqref{newME} has recently been used to model a double quantum dot setup exhibiting \PT~symmetry~\cite{PurkayasthaPRR2020}. Hence, we set parameter values matching those considered in Ref.~\cite{PurkayasthaPRR2020}: $g=2$ (MHz), $\gamma_G=0.6g$ and $\Gamma=1.16g$, so to ensure that $\gamma_L^\T{PT}<\gamma_L^\T{EP}<\gamma_L^\T{th}$.

\begin{figure}
			\includegraphics[width=0.95\linewidth]{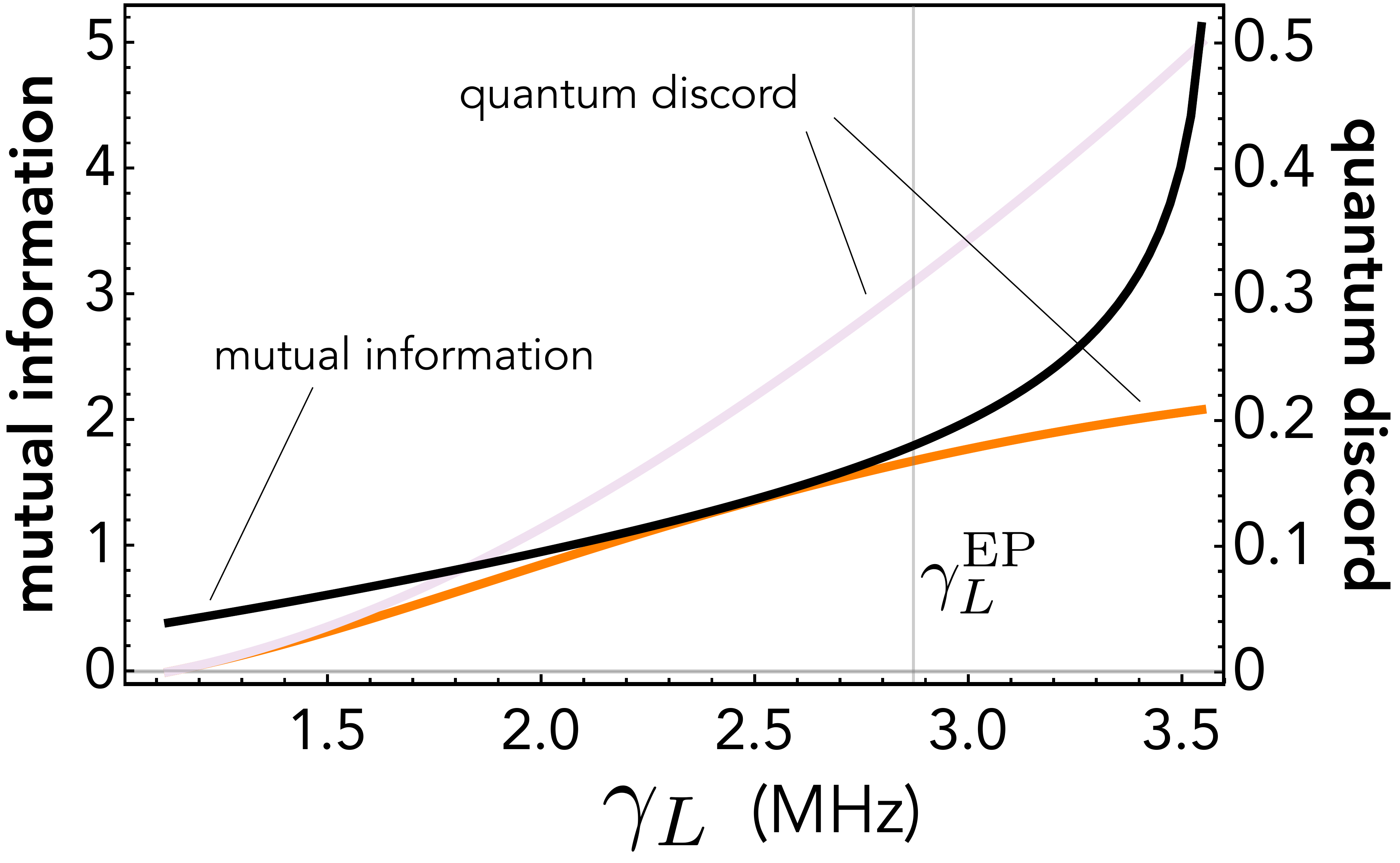}
			\caption{
				{\it Asymptotic correlations in the non-\PT~regime}. 
				Stationary value of mutual information  [left scale] in black and quantum discord [right scale] $\mathcal D_{GL}$ and $\mathcal D_{LG}$ in pink and orange, respectively. Contrarily to the \PT-symmetric regime, they both appear to be insensitive to the EP occurring at $\gamma_L=\gamma_{L}^{\rm EP}$.
			} \label{fig7}
		\end{figure}

\begin{figure}
			\includegraphics[width=\linewidth]{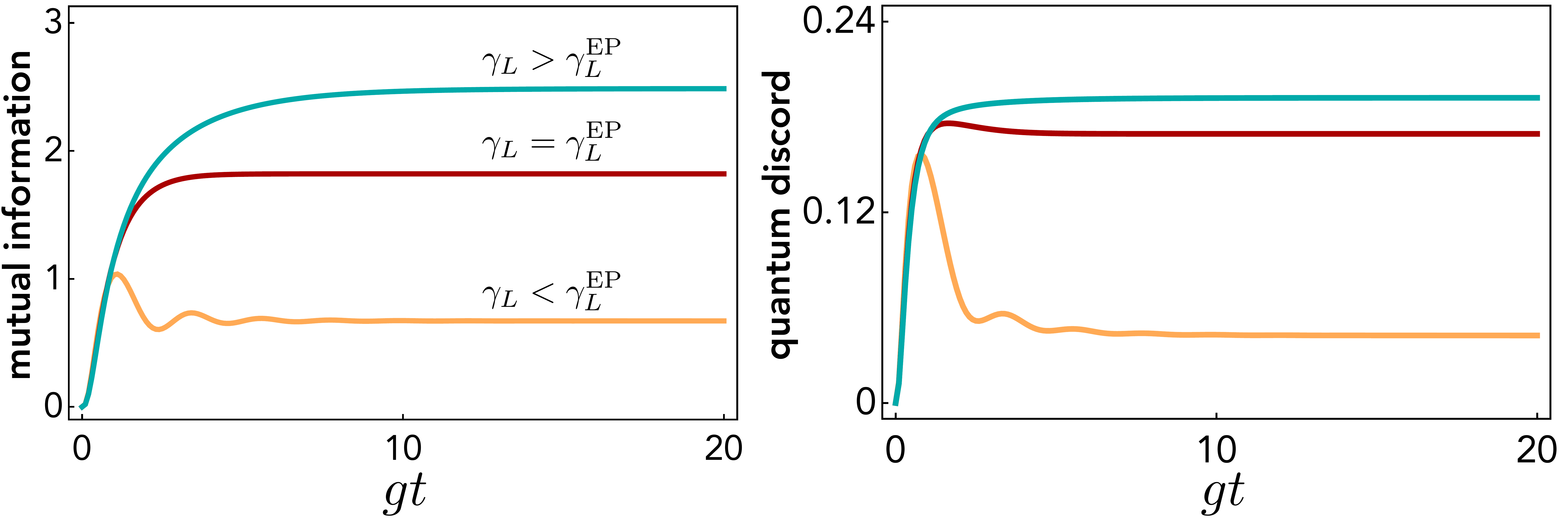}
			\caption{
				{\it Total and quantum correlations dynamics in the non-\PT~regime}. 
				Time behavior of mutual information (left) and quantum discord $\mathcal D_{LG}$ (right)  below 
				[$\gamma_L=0.8g$],
				at and above 
				[$\gamma_L=1.6g$] 
				the EP (orange, red and cyan, respectively). Above the EP, oscillations are overdamped.
			} \label{fig8}
		\end{figure}

		
		Contrarily to the \PT-symmetric regime, total and quantum correlations are not sensitive at large times to this EP. As displayed in Fig.~\ref{fig7} quantum correlations can amount up to 10\% of total ones right at the lasing threshold, where they also achieve half of the entanglement threshold (see Appendix~\ref{appendix}). 
		A different behavior below and above the EP  can be spotted only by looking at the transient dynamics. Indeed, as shown in Fig.~\ref{fig8}, below (above) the EP both mutual information and discord (do not) exhibit oscillatory behavior before reaching their stationary value.

		\section{Conclusions}

		We investigated the behavior of quantum and total correlations in a \PT-symmetric system beyond the pure balanced gain and loss picture. 
		
		We first studied the effect of adding dissipation to both modes of a gain-loss setup, which still enjoys PT~symmetry. We found that the qualitative time behavior of mutual information and quantum discord in this \textit{dissipative} gain-loss system is similar to the pure gain-loss case. In particular, total correlations increase, while quantum ones are reduced. Also the asymptotic behavior of QCs is unchanged,  vanishing in the unbroken regime and being finite in the broken one.
		
		We made a similar investigation across an EP without \PT~symmetry, which appears by tuning the loss strength. In this case we still observe different dynamical behaviors below, at and beyond the EP, though these differences disappear at large times.
		
		We observe how asymptotic distinct behaviors across the EP occur only in presence of \PT~symmetry. This suggests a deeper interplay between this symmetry and long-time properties, which is left for future investigation.

		\section*{Acknowledgments}
		
		We acknowledge support from MIUR through project PRIN Project 2017SRN-BRK QUSHIP.
	
		\appendix
		
		\section{Classical and quantum correlations}\label{appendix}

		In order to compare our results to those of~\cite{roccatiQST2021}, we focus on two quantities: mutual information, which captures the total amount of correlations among the two modes, and quantum discord, which by definition measures the pure quantum correlation between them.
		The former is defined as $\mathcal I= S_{G}+S_{L}-S$ where $S_{L(G)}=-\T{Tr}(\rho_{L(G)}\log\rho_{L(G)})$ and $S=-\T{Tr}(\rho\log\rho)$~\cite{cover2006,nielsen2010} are the local and global entropies, respectively. Mutual information is zero only for product states, therefore it captures the \textit{total} amount of correlations. Quantum discord instead is defined by the difference of total and classical correlations~\cite{ollivierPRL2001,hendersonJPAMG2001,modiRMP2012} and can be expressed as

		\begin{equation}
		\mathcal D_{LG}=S_{G}-S+\underset{\hat G_k}{\rm min}\,\,\sum_k p_k S(\rho_{L|k})\,,\label{d-def}
		\end{equation}

		where $\rho_{L|k}=\Tr_G(\hat G_k \rho)/p_k$ is the state after a measurement $\hat G_k$ made on $G$ with outcome $k$ and the minimization is over all possible measurements.
		Similarly, $\mathcal D_{GL}$ is obtained by swapping $G$ and $L$ in \eq\eqref{d-def}.
		More explicitly, the total von Neumann entropy is given by $S=f(\nu_-)+f(\nu_+)$, 
		where $f(x)=\frac{x+1}{2}\log[\frac{x+1}{2}]-\frac{x-1}{2}\log[\frac{x-1}{2}]$, $\nu_\pm$ are the symplectic eigenvalues given by $2\nu_\pm^2=\Delta\pm\sqrt{\Delta^2-4|\sigma|}$, $\Delta=|A|+|B|+2|C|$, the two-mode covariance matrix is
		\begin{equation}
            \sigma = 
            \begin{pmatrix} 
            L & C \\ 
            C^\text{T} & G 
            \end{pmatrix},
        \end{equation}		
		and $|M|=\det M$. The local entropies are instead $S_{M}=f(|M|)$, with $M=G,L$. The explicit form of quantum discord is
		$\mathcal D_{LG}
		    =
		    f(\sqrt{|G|})
		    -
		    f(\nu_-)
		    -
		    f(\nu_+)
		    +
		    \tilde f$
		\begin{equation}
		    \tilde f
		    =
		    \begin{cases}
		            \frac{2|C|^2+(|G|-1)(|\sigma|-|L|)+2 \abs{|C|} \sqrt{|C|^2+(|G|-1)(|\sigma|-|L|)}}{(|G|-1)^2}\\
		   \qquad\qquad\qquad\qquad\qquad\qquad\qquad\qquad\text{if } \delta <0,
		            \\
		            \frac{|G||L|-|C|^2 +|\sigma| -
		            \sqrt{|C|^4 + (|\sigma|-|G||L|)^2-2|C|^2(|\sigma|+|G||L|)}
		            }{2|G|}
		            \\
		 \qquad\qquad\qquad\qquad\qquad\qquad\qquad\qquad\text{otherwise.}
		    \end{cases}
		\end{equation}
		where $\delta=(|\sigma|-|G||L|)^2-|C|^2(|G|+1)(|\sigma|+|L|)$.
		The analogous expression for $\mathcal D_{GL}$ is obtained by exchanging $G$ with $L$.
		
		Notice that discord is generally asymmetric, i.e., $\mathcal D_{LG}\neq \mathcal D_{GL}$.
		{Based on its definition,} discord captures quantum correlations beyond entanglement, as  separable states are generally discordant~\cite{modiRMP2012}. {Therefore, the two quantum oscillators are classically correlated only when $\mathcal D_{LG(GL)}=0$ and ${\cal I}\neq 0$.}

		In the special case of Gaussian states, the minimization in~\eqref{d-def} can be restricted to Gaussian
		measurements~\cite{pirandolaPRL2014}, which yields an explicit expression for \textit{Gaussian} discord ${\mathcal D}$~\cite{giordaPRL2010,adessoPRL2010} capturing correlations beyond entanglement since  states with $\mathcal D>1$ are entangled~\cite{adessoPRL2010}.

		\bibliographystyle{apsrev4-1}			
		\bibliography{biblio}

\end{document}